# Testing sub-gravitational forces on atoms from a miniature, in-vacuum source mass

Matt Jaffe[1*], Philipp Haslinger[1*], Victoria Xu[1], Paul Hamilton[2], Amol Upadhye[3],
Benjamin Elder[4], Justin Khoury[4], Holger Müller[1,5]

[1]Department of Physics, 366 Le Conte Hall MS 7300, University of California, Berkeley, California 94720, USA.
[2]Department of Physics and Astronomy, University of California, Los Angeles, California 90095, USA
[3]Department of Physics, University of Wisconsin-Madison, Madison, WI 53706, USA.
[4]Center for Particle Cosmology, Department of Physics and Astronomy, University of Pennsylvania, Philadelphia, PA 19104, USA.
[5]Lawrence Berkeley National Laboratory, One Cyclotron Road, Berkeley, CA, 94720, USA.
*These authors contributed equally to this work

Gravity is the weakest fundamental interaction and the only one that has not been measured at the particle level. Traditional experimental methods, from astronomical observations to torsion balances, use macroscopic masses to both source and probe gravitational fields[1]. Matter wave interferometers[2] have used neutrons[3], atoms[4] and molecular clusters[5] as microscopic test particles, but initially probed the field sourced by the entire earth. Later, the gravitational field arising from hundreds of kilograms of artificial source masses was measured with atom interferometry[6,7]. Miniaturizing the source mass and moving it into the vacuum chamber could improve positioning accuracy, allow the use of monocrystalline source masses for improved gravitational measurements, and test new physics, such as beyond-standard-model ("fifth") forces of nature and non-classical effects of gravity. In this work, we detect the gravitational force between freely falling cesium atoms and an in-vacuum, centimeter-sized source mass using atom interferometry with state-of-the-art sensitivity. The ability to sense gravitational-strength coupling is conjectured to access a natural lower bound for fundamental forces[8], thereby representing an important milestone in searches for physics beyond the standard model. A local, in-vacuum source mass is particularly sensitive to a wide class of interactions whose effects would otherwise be suppressed beyond detectability[9,10] in regions of high matter density. For example, our measurement strengthens limits on a number of cosmologically-motivated scalar field models, such as chameleon[10] and symmetron fields[11,12], by over two orders of magnitude and paves the way toward novel measurements of Newton's gravitational constant $G$ and the gravitational Aharonov-Bohm effect[13].

We measure the acceleration experienced by atoms near a miniature, in-vacuum source mass using light-pulse atom interferometry[4]. This technique is based on the wave-particle duality of quantum mechanics and transduces the acceleration experienced by atoms into a phase difference between interfering atomic matter-waves. In our setup, cesium atoms are laser-cooled and launched[14] upwards into free fall. Pulses from counter-propagating laser beams transfer them from their initial quantum state $|a\rangle$ to another state $|b\rangle$. Each atom absorbs one photon, having a momentum $\hbar k_1$, from the first beam while simultaneously being stimulated to emit another photon into the second beam, gaining additional momentum $\hbar k_2$. This results in a total momentum change of $\hbar k_{\text{eff}}$, where $k_{\text{eff}}=k_1+k_2$. The first interferometer pulse has a duration such that the transfer takes place with 50% probability. It acts as a coherent beam splitter for matter waves, placing the atom into a superposition of the initial state $|a, p_0\rangle$ with momentum $p_0$ and the state $|b, p_0+\hbar k_{\text{eff}}\rangle$. The two states separate spatially. After a pulse separation time $T$, a second laser pulse transfers the states $|a, p_0\rangle \rightarrow |b, p_0+\hbar k_{\text{eff}}\rangle$ and $|b, p_0+\hbar k_{\text{eff}}\rangle \rightarrow |a, p_0\rangle$, thus inverting the relative motion. After another interval $T$, a third pulse acts as a final beam splitter which combines the partial matter waves (Fig. 1). When combined, the matter waves add constructively or destructively, depending on their phase difference $\Delta\phi$, giving a probability $P\sim\cos^2(\Delta\phi/2)$ of finding the atom in the quantum state $|a\rangle$. The probability and thus the phase difference is found by measuring the population ratio of states $|a\rangle$ and $|b\rangle$ when many atoms undergo this process simultaneously. In the simplest case, $\Delta\phi = k_{\text{eff}} aT^2$, where $a$ is the acceleration of the atoms. Using $k_{\text{eff}} \sim 10^7$/m, large atom samples for sufficient phase resolution, and $T$ of up to 1.15 seconds[15] creates an enormous lever arm by which even small changes of $a$ generate measurable phase changes.

Unfortunately, this lever arm is significantly shortened when it comes to measuring the gravitational force created by a small mass $M$. The useful free-fall time, distance between the atoms and the source mass, and the dimensions of the atom cloud are all constrained because atoms far away from the source see a reduced gravitational potential. Accordingly, the most sensitive atom interferometers[16] use the entire earth as a source mass. Measurements using small source masses have not been previously demonstrated, but could be useful to explore new regimes[13,17].

Testing gravity in new ways may help to answer pressing questions. Cosmological measurements[18,19] have firmly established that the universe is expanding at an accelerating rate which is consistent with dark energy permeating all of space. The observed dark energy density $\Lambda_0^4 \approx (2.4\text{ meV})^4$ is tens of orders of magnitude smaller than expected from the vacuum energy of quantum field theories. This chasm, the "cosmological constant problem," likely requires new fields for its resolution. By Weinberg's no-go theorem[20], however, even such new fields cannot solve the problem unless they are dynamic, not in equilibrium. A new field must therefore be light if it is to address the cosmological constant problem, ($m \leq H_0$ where $H_0 \sim 10^{-33}$ eV is the Hubble constant) so as to remain in non-equilibrium today, $10^{10}$ years after the Big Bang. Such a light field, however, should mediate a long-range interaction, in disagreement with precision tests of gravity. Over the last decade, this has motivated a family of theories that predict significant deviations from general relativity only in the *ultra weak-field* regime[20], where a force of gravitational strength or larger is suppressed further by "screening" as a function of the environment. Existing theories do not solve the cosmological constant problem, but screening is likely a key ingredient of any future solution. Experimental tests of gravity have focused on the short-distance regime, the post-Newtonian regime or the strong field regime[21], leaving the ultra-weak field regime largely untested.

Screening arises when coupling between the field and matter hides effects of the field (such as a new force) in high-density regions like Earth. In contrast, the field is unsuppressed and most potent in low-density regions like the cosmos. Such fields can enact important astrophysical effects while remaining hidden from laboratory and solar system tests.

The ultra-weak fields $\varphi$ can be characterized by their mass $m(\varphi)$ and coupling to normal matter $\beta(\varphi)$, which may both be functions of the field itself. The acceleration of an object

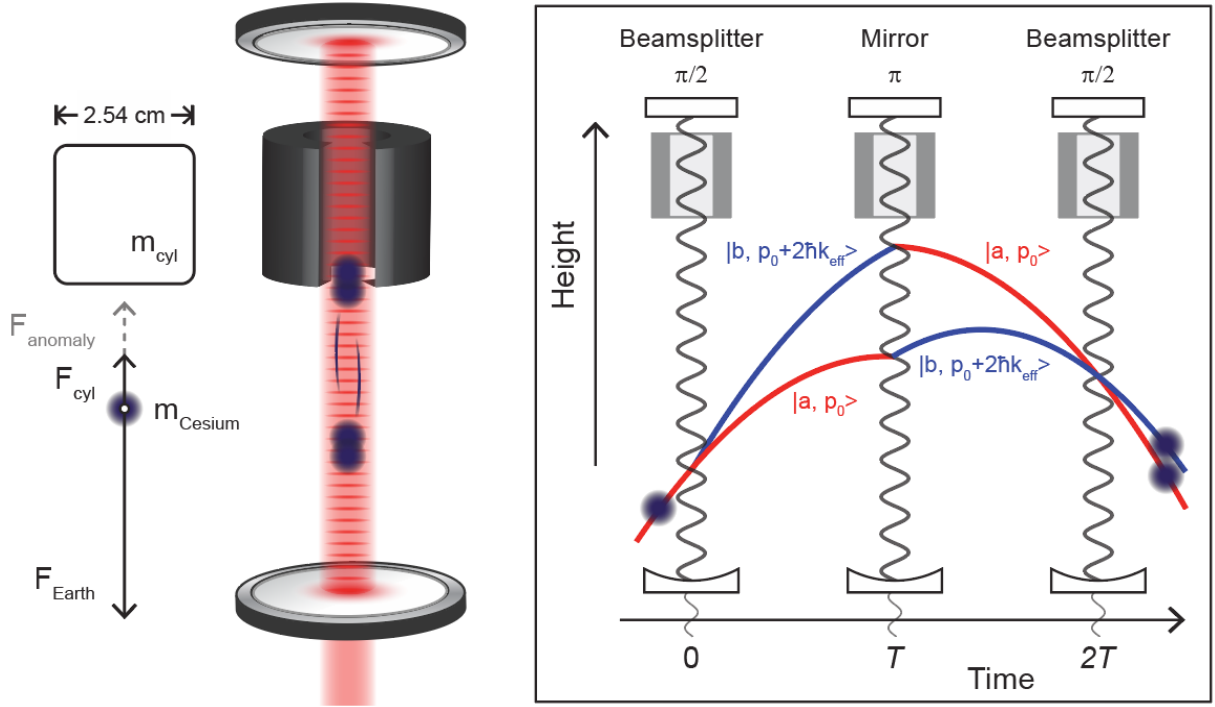

**Figure 1 | Cavity matter-wave interferometry.** Left: Experimental setup. The acceleration $a_{cyl}$ of cesium atoms towards a cylindrical tungsten source mass suspended in ultra-high vacuum is measured. The cylinder has mass $m_{cyl}$ = 0.19 kg, height $h$ and diameter $d = h = 2.54$ cm. The axial through-hole has radius 0.5 cm, and the slot has width 0.5 cm. Making a differential measurement isolates the effect of any interactions sourced by the tungsten mass. Right: Mach-Zehnder interferometer based on Raman transitions in an optical cavity. Three laser pulses manipulate the cesium atoms during free-fall. The pulses 1) split the atomic wave packet along two different trajectories, 2) reflect the two trajectories near their apex, and 3) recombine and interfere the matter waves to measure the phase difference accumulated between the two paths during the interferometer time of $2T = 110$ ms. We obtain a measurement of the acceleration experienced by the cesium atoms ensemble-averaged over ~$10^5$ atoms.

$$a = -\frac{\beta(\varphi)}{M_{Pl}} \lambda_a \nabla \varphi \quad (1)$$

(in our case, an atom) caused by the field is highly sensitive to the surrounding matter geometry[22]. Here, $M_{Pl}=(\hbar c/8\pi G)^{1/2} \approx 2.4 \times 10^{18}$ GeV is the reduced Planck mass, and $0 \leq \lambda_a \leq 1$ is a screening function that depends on $m$, $\beta$ and the object's mass and size. Moreover, $\lambda_a \to 1$ for a small and light test particle but $\lambda_a \ll 1$ for macroscopic objects, where only a thin, outermost layer interacts with the field. An atom in ultra-high vacuum with a local miniature source mass minimizes screening and is well-suited as a test mass for such theories[9]. Prime examples of such scalar fields are chameleons and symmetrons.

A **chameleon** scalar field[23,24] is characterized by an effective potential density

$$V_{eff}(\varphi) = V(\varphi) + V_{int}(\varphi). \quad (2)$$

The self-interaction

$$V(\varphi) = \Lambda^4 + \frac{\Lambda^{4+n}}{\varphi^n} \quad (3)$$

is characterized by an energy scale $\Lambda$, which must be close to the cosmological-constant scale, $\Lambda \simeq \Lambda_0 = 2.4$ meV, if the chameleon is to drive cosmic acceleration. The interaction with matter of density $\rho_m$

$$V_{int} = \rho_m \varphi/M \quad (4)$$

is characterized by an energy scale $M$, which is expected to be below the Planck mass. The chameleon profile due to an arbitrary static distribution of matter $\rho_m(\vec{x})$ is obtained by solving the non-linear Poisson equation:

$$\nabla^2 \varphi = \partial V_{eff}/\partial \varphi. \quad (5)$$

Deep inside a large, dense object, $\nabla^2 \varphi \simeq 0$ and $\varphi$ rapidly approaches a negligible value that minimizes $V_{eff}(\varphi)$. Thus, the bulk of such an object is largely decoupled from the field, except for a thin outer shell, leading to screening. For general $\rho_m(\vec{x})$, we must resort to numerical integration[22]. Given the resulting field profile $\varphi(\vec{x})$, the chameleon-mediated acceleration on an atom is given by equation (1) with $\beta_{cham}=M_{Pl}/M$.

A **symmetron** scalar field[25,26] has an effective potential symmetric under $\varphi \to -\varphi$. The simplest models have a Higgs-like quartic self-interaction

$$V(\varphi) = \frac{\lambda}{4!}\varphi^4 - \frac{\mu^2}{2}\varphi^2, \quad (6)$$

in which $\lambda$ is the self-coupling, and $\mu$ is the bare potential mass scale. The field couples to matter through an explicitly density-dependent mass term,

$$V_{int}(\varphi) = \frac{\rho_m}{2M_S^2} \varphi^2. \quad (7)$$

The coupling is again characterized by an energy scale $M_S$. We focus here on 1 MeV < $M_S$ < 1 TeV, approximately the regime in which the fifth force is screened in a typical laboratory apparatus. The acceleration equation (1) in a constant-density region is roughly characterized by $\beta_{sym}(\varphi) = \varphi M_{Pl}/M_S^2$. The field $\varphi$, and thus the coupling $\beta_{sym}$, is zero in high-density regions and nonzero at low densities. A sharp transition away from the symmetric, uncoupled phase will only occur in a vacuum chamber larger than $\pi/\mu$, and forces are suppressed at distances much smaller than $1/\mu$, so the range[11] of $\mu$ probed by our experiment is approximately 0.01 meV < $\mu$ < 1 meV.

Our basic setup has been described previously[10]. In order to reach the sensitivity to observe the gravitational attraction between the atoms and the source mass, we installed significant technical upgrades. 3D Raman sideband cooling reduces the atom temperature to ~300 nK. Launching the atoms vertically upwards in a cavity-enhanced optical lattice doubles the available interrogation time and quadruples the accumulated phase. Two levels of passive vibration isolation attenuate seismic noise, and an active stabilization loop provides further attenuation. After the interferometer, performing the lattice launch in reverse "catches" atoms remaining in the cavity mode while the rest fall away. This spatial selection of atoms participating in the interferometer increases contrast by an order of magnitude to over 40%.

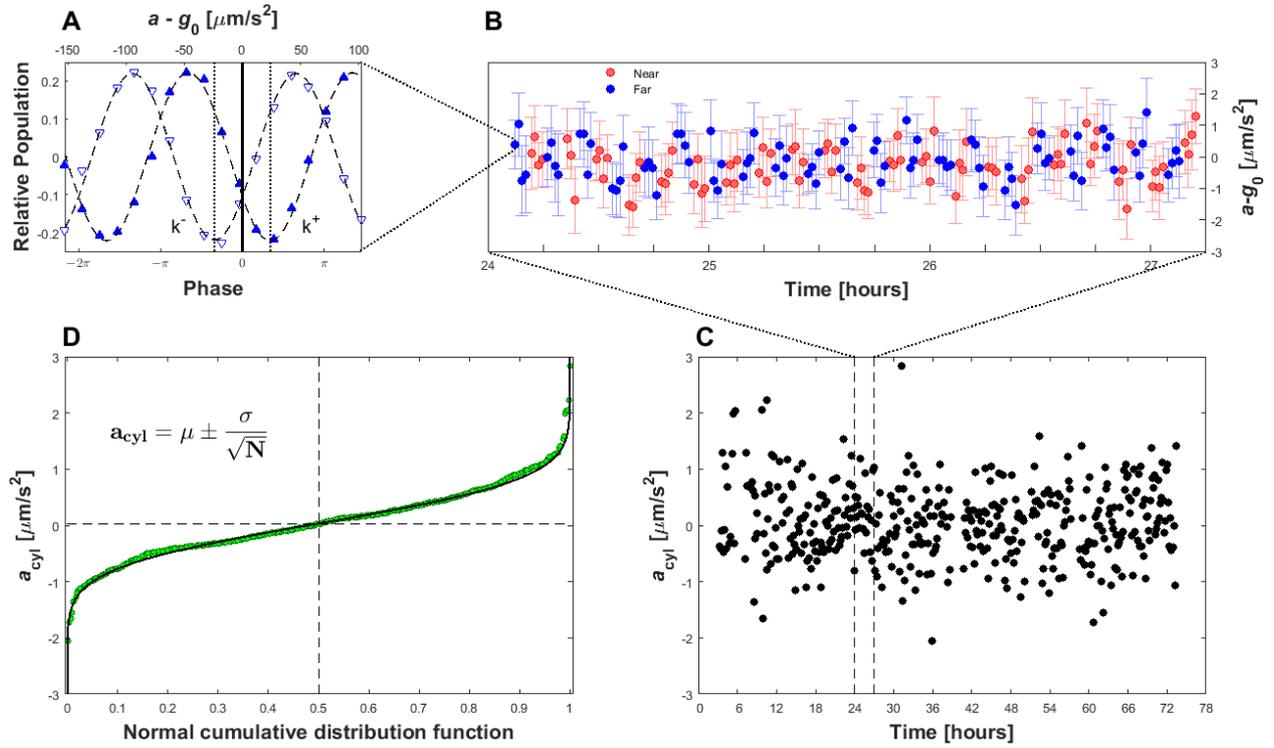

**Figure 2 | Experimental data.** A differential measurement is performed by toggling the source mass between a near and far position. **A)** Two interferometer fringes, taken for the wavevector direction, up ($k^+$) and down ($k^-$) (see Methods). Combining these two measurements gives one data point out of 3215 taken over 68 hours. **B)** A 3-hour section of data. Four measurements are taken at each source mass position, which are then averaged. Blue points indicate that the source mass is in the far position; red points indicate the near position. An overall offset $g_0$ is subtracted for clarity. The difference between subsequent measurements after toggling the source mass position gives one measurement of $a_{cyl}$. **C)** A single determination of $a_{cyl}$ takes ~500 seconds. The full dataset (one of three) is shown here. **D)** The set of individual $a_{cyl}$ measurements are fitted to the cumulative distribution function of a normal distribution with mean $\mu$ and standard deviation $\sigma$. $N$ is the number of individual measurements. This least-squares fit (solid black line) gives $a_{cyl}$ for the dataset.

A schematic of our apparatus is shown in Fig. 1. Laser beams inside a Fabry-Pérot cavity provide well-controlled optical wavefronts and resonant power enhancement for coherent manipulation of the atomic probe. A tungsten cylinder of mass $m_{cyl} = 0.19$ kg is our source mass. We have optimized this geometry using detailed numerical calculations of screened field profiles[22] in our vacuum chamber. An axial through-hole allows the cavity mode to pass through the mass unimpeded. A rectangular slot along one side of the cylinder allows the mass to be toggled between one position *near* the atoms and another position *far* away, without interrupting the cavity mode. A differential measurement between the two positions suppresses the earth's gravitational acceleration and isolates the acceleration arising from the cylindrical source mass, $a_{cyl} = a_{near} - a_{far}$. This acceleration should be purely gravitational in the absence of any anomalous interactions.

Data was taken for more than 170 hours through three quiet weekends in Oct. 2016, resulting in ~$4.3 \times 10^5$ experimental runs (see Fig. 2). Averaging the measurements of the acceleration $a_{cyl}$ weighted by the standard error over these 3 datasets results in $a_{cyl} = (76 \pm 19_{stat} \pm 16_{syst})$ nm/s$^2$, where the first error bar (one standard deviation) is statistical and the second arises from systematic uncertainties (see Methods). The positive acceleration indicates a force toward the source mass. This agrees well with the expected gravitational pull of the cylinder $a_{grav} = (65 \pm 5)$ nm/s$^2$. We obtain an anomalous acceleration $a_{anomaly} = a_{cyl} - a_{grav} = (11 \pm 24)$ nm/s$^2$, giving a 95% confidence interval of -37 nm/s$^2$ < $a_{anomaly}$ < 59 nm/s$^2$. Using a one-tailed test to bound fifth force interactions (which must be attractive for scalar fields with a universal matter coupling), we constrain anomalous accelerations $a_{anomaly}$ < 50 nm/s$^2$ at the 95% confidence level. We note that the 24 nm/s$^2$ = 2.4 ng (1σ) accuracy is on par with the most accurate atom interferometric gravity measurements (e.g., 7 nm/s$^2$ in [4]) and the sensitivity was achieved within a free-fall distance of 1.4 cm and an atom cloud within a 600-μm radius interferometer beam; when scaled for these dimensions, our miniature interferometer compares favorably with large-scale ones.

Specializing to chameleon and symmetron fields, following Burrage et al.[9], we improve previous limits[10,11] on these models by more than 2 orders of magnitude. Fig 3 shows excluded parameter ranges for these models. For chameleon fields with $\Lambda$ at the dark energy value $\Lambda_0 = 2.4$ meV and $n = 1$, we exclude up to $M < 2.7 \times 10^{-3} M_{Pl}$, narrowing the gap to torsion pendulum constraints[1,27]. One can see that these fields are nearly ruled out, with only a one order of magnitude range left for the coupling strength $M$. Furthermore, for all $\Lambda > 5.1$ meV, this gap is fully closed, ruling out all such models. Our symmetron limits are complementary to torsion pendula[1,11] as well. We improve previous constraints on $\lambda$ by two orders of magnitude throughout the entire range of $M_S$ and $\mu$ probed by our experiment. Our constraint is strongest in the regime where the atom is screened, where for $\mu = 0.1$ meV we rule out $\lambda < 1$.

Tests of gravity in the ultra-weak field regime with a miniature, in-vacuum source mass probed screened field theories with the potential to explain the accelerated expansion of our universe. In the future, technologies such as lattice interferometry[14] in our optical cavity and large momentum transfer Bragg beam splitters will allow us to hold quantum probe particles in proximity to a miniature source mass, evading geometric constraints from the source mass' small size, and boosting sensitivity. With modest improvements, chameleon fields at the cosmological energy density will be either discovered or completely ruled out. This also will enable study of novel quantum phenomena such as the gravitational Aharonov-Bohm effect[13], and provide even better resolution of atom – source mass interaction.

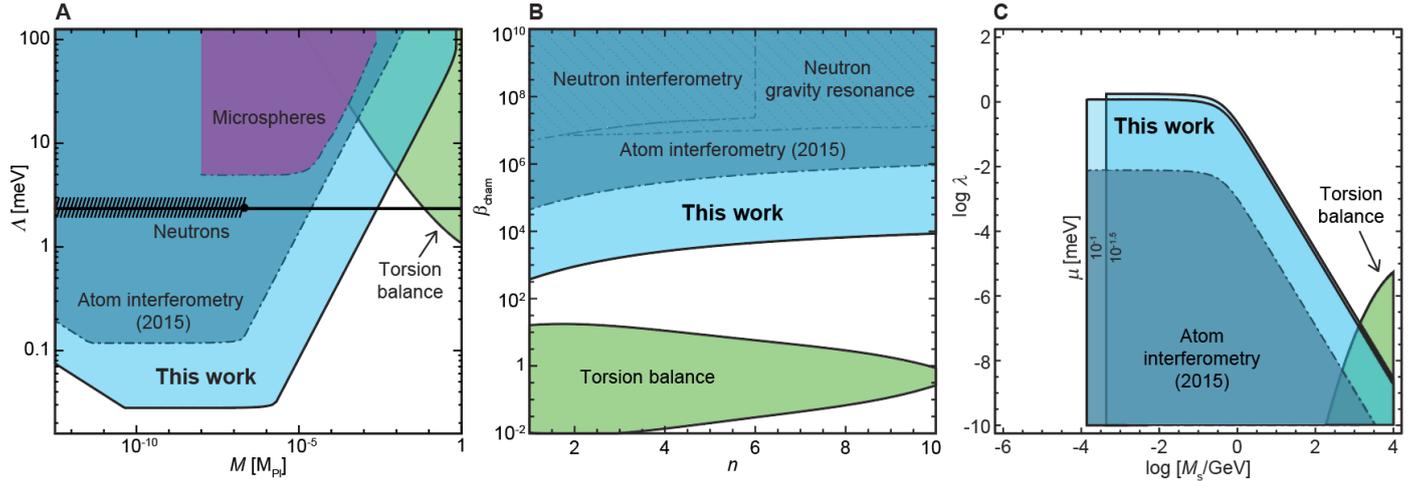

**Figure 3 | Constraints on screened scalar fields. A)** Chameleon field: The shaded areas in the $M$-$\Lambda$ plane are ruled out at the 95% confidence level. $M_{Pl}/M$ gives the coupling strength to normal matter in relation to gravity; $\Lambda = \Lambda_0 \approx 2.4$ meV (indicated by the black line) could drive cosmic acceleration today. A comparison is made to previous experiments: neutron interferometry[28] / neutron gravity resonance[29], microsphere force sensing[30], and torsion balance[1,27]. **B)** Chameleon limits in the $n$-$\beta_{cham}$ plane with $\Lambda=\Lambda_0$, showing the narrowing gap in which basic chameleon theories could remain viable. $n$ is the power law index describing the shape of the chameleon potential; $\beta_{cham} \equiv M_{Pl}/M$ is the strength of the matter coupling. **C)** Symmetron fields: Constraints by atom interferometry complement those from torsion pendulum experiments[11] (shown with $\mu = 0.1$ meV) for the range of $\mu$ considered. For $\mu < 10^{-1.5}$ meV, the field vanishes entirely inside the vacuum (see Methods), leaving this parameter space unconstrained. The same effect produces the sharp cutoff in our limits at low $M_S$.

**Acknowledgements**
We thank Brian Estey for helpful discussions and technical contributions to the apparatus. This material is based upon work supported by the National Science Foundation under CAREER Grant No. PHY-1056620, the David and Lucile Packard Foundation, and National Aeronautics and Space Administration Grants No. 1553641, No. 1531033, and No. 1465360. We also acknowledge collaboration with Honeywell Aerospace under DARPA Contract No. N66001-12-1-4232. P. Has. thanks the Austrian Science Fund (FWF): J3680. B.E. and J.K. are supported in part by NSF CAREER Award PHY-1145525, NASA ATP grant NNX11AI95G, and the Charles E. Kaufman Foundation of the Pittsburgh Foundation.

**Author Contributions**
M.J., P.Has., V.X., P.Ham. and H.M. built the apparatus, took the measurements, and performed the data analysis. B.E. performed numerical simulations of screened fields. A.U., B.E. and J.K. interpreted the measurements in the context of screened fields. All authors contributed to the manuscript.

**Author Information**
Correspondence and requests for materials should be addressed to H.M. (hm@berkeley.edu).

# METHODS

## Setup

We describe the basic outline and new features of our setup[10,31]. Cesium atoms are loaded into a three-dimensional magneto-optical trap (3D-MOT) from a 2D-MOT. After sub-Doppler cooling in an optical molasses, we perform Raman sideband cooling[32] in a 3D lattice which leaves $\sim 5\times 10^6$ atoms in the $|F = 3, m_F = 3\rangle$ state at a temperature < 300 nK. After release from the lattice, adiabatic rapid passage and a state selection pulse with microwaves transfer the cold atoms into the magnetically insensitive $|F = 3, m_F = 0\rangle$ state. About 20% of the atoms are then launched upwards with a chirped optical lattice[14] in the optical cavity mode at a velocity of 59.1 cm/s. Launching the atoms moves them upwards towards the source mass, doubles the available interrogation time, and provides both spatial and velocity selection. After the launch, we perform the interferometry pulse sequence. The cavity dictates that all beams counter-propagate. Close to the apex, the Doppler shift $\delta_{\text{Dopp}}$ due to atom motion is small. The frequencies driving Raman transitions imparting upward momentum ($k^+$) therefore become degenerate with the ones imparting downward momentum ($k^-$). Since this would cause atom loss, we make our interferometer asymmetric with respect to the apex of the atomic trajectory. In order to increase signal to noise, we preferentially detect atoms at the center of the Raman beam, suppressing the signal from atoms that have not participated in the interferometer. To this end, after the interferometer we reverse the launch procedure in order to catch the atoms by decelerating them into a lattice at zero velocity. This selects only the atoms in the center of the cavity mode, while nonparticipating atoms (*e.g.*, that have left the cavity mode due to thermal expansion) fall away. A pushing beam separates the two output ports of the interferometer, where they are counted by fluorescence detection to determine their relative population.

## Vibration isolation

Vibrations of the retroreflecting cavity mirrors are a leading noise source. We mount the entire vacuum chamber on two layers of passive vibration isolation: a pneumatic benchtop isolation system (Thorlabs PWA 090), which is mounted on top of a floated optical table. For active isolation, a seismometer (Kinemetrics Inc., SS-1) sits on top of the vacuum chamber to measure residual vibrations. The seismometer signal enters an analog feedback loop which actively stabilizes against vibrations using a voice coil actuator. The seismometer is magnetically shielded from switching experimental magnetic fields with a cylindrical pipe of low-carbon steel, reducing synchronous accelerations induced by the servo loop.

The passive isolation attenuates ground vibrations by up to two orders of magnitude. Closing the servo loop reduces the seismometer error signal a further factor of ~5 - 400 from 1-20Hz, the most problematic frequency range for our $2T = 110$ ms interferometer. This active servo loop reduces the expected interferometer phase noise by a factor of 16.

## Systematic effects

Taking the difference of measurements with the source mass *near* and *far* from the interferometer cancels systematic phase shifts that are independent of the source mass position. Such effects include Earth's gravity and gravity gradients. Also included are deviations of the phase of our laser beam from that of a hypothetical perfect plane wave, *e.g.*, from the Gouy phase and wavefront curvature. The cavity mode ensures retroreflection alignment (*i.e.*, that $\vec{k}_1$ and $\vec{k}_2$ are anti-parallel). The remaining systematic effects are discussed below, and summarized in Extended Data Table 1.

Many systematics can be suppressed by wavevector reversal. If $\vec{k} \to -\vec{k}$ (*i.e.*, the atoms are kicked down ($k^-$) rather than up ($k^+$)) the sign of the acceleration phase $\Delta\phi_{\text{acc}} = \vec{k}_{\text{eff}} \cdot \vec{a}\, T^2$ changes but certain systematic phases, such as Zeeman shifts and ac Stark shifts (to first order) do not. We can invert the sign of the effective laser wavevector ("wavevector or k-reversal") by changing the frequency difference of the Raman beams, the so-called Doppler detuning $\delta_{\text{Dopp}}$. Averaging acceleration measurements for both $k^+$ and $k^-$ allows us to subtract out Zeeman and ac Stark phase shifts, leaving only the acceleration phase.

### *Zeeman shift*
A Zeeman shift dependent on the source mass position could cause a phase shift mimicking an acceleration. Zeeman shifts can enter into the interferometer phase through the cesium hyperfine energy splitting $\alpha_B = 2\pi \cdot 427.45$ Hz/G$^2$ in a magnetic field, though only quadratically since we are using first-order magnetically insensitive states. The Zeeman phase $\Delta\phi_{\text{Zeeman}}$ is calculated by integrating the classical action along the trajectory of the two interferometer arms.

The tungsten source mass should be non-ferromagnetic, but impurities or eddy currents could still cause small magnetic fields. We therefore measure the magnetic field along the trajectory for both positions of the source mass, with the same experimental timing, so that MOT eddy currents are included. To do so, we measure the magnetically-sensitive $|F = 3, m_F = 3\rangle \to |F = 4, m_F = 4\rangle$ microwave using the atoms as a local probe. The individual measurements vary with source mass position only by ~0.1%. We fit the field measurements to a quadratic $B = B_0 + B_1 z + B_2 z^2$, where $z$ is a spatial coordinate, as in Extended Data Fig. 1. This gives the magnetic field parameters $B_0$, $B_1$, and $B_2$, which are used to calculate $\Delta\phi_{\text{Zeeman}}$.

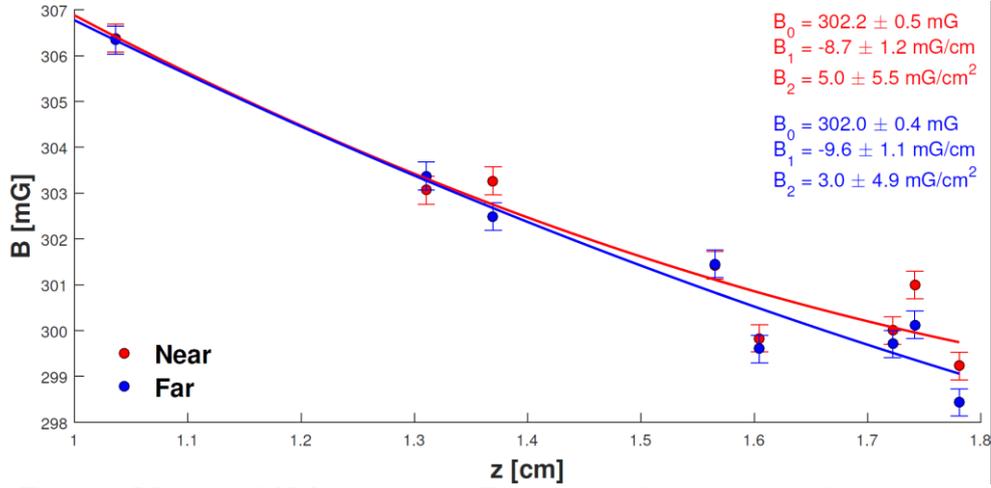

**Extended Data Figure 1 | Magnetic field determination.** The magnetic field is measured for both positions of the source mass along the atomic trajectory.

We can calculate $\Delta\phi_{\text{Zeeman}}$ for our atomic trajectory in the measured field for both wavevector directions $k^{\pm}$. The expression for $\Delta\phi_{\text{Zeeman}}$ is non-reversing under the transformation $k \to -k$, unlike the acceleration phase $\Delta\phi_{\text{acc}}$ that we are interested in. Cancellation is imperfect, however, because the k-reversed interferometers are kicked in opposite directions, leading to slightly different classical trajectories. We cancel >90% of $\Delta\phi_{\text{Zeeman}}$ using k-reversal.

The difference in Zeeman phase ($\Delta\phi_{\text{Zeeman}}^{\text{near}} - \Delta\phi_{\text{Zeeman}}^{\text{far}}$) after k-reversal is 580 µrad corresponding to a 13 nm/s$^2$ shift in the measured acceleration. This is the same shift we calculate by using numerical integration of a linear interpolation function between the measured points, confirming that the quadratic fit captures the details of the magnetic field. Furthermore, it's likely that most of the measured difference actually arises from variation in the Rabi frequency of the microwaves used to probe the transition, as similar frequency shifts were seen in the less magnetically-sensitive $|F = 3, m_F = 0\rangle \to |F = 4, m_F = 1\rangle$ transition.

*AC Stark shift*
The ac Stark shift causes differential energy shift of the cesium $F = 3$ and $F = 4$ levels in a light field. This shift is opposite during interferometer pulses 1 and 3 (when the arms are in different states), which leads to a phase difference. This phase cancels in the symmetric Mach-Zehnder geometry if pulses 1 and 3 are identical. However, asymmetry between these pulses (due to thermal expansion of the cloud, changing $\delta_{\text{Dopp}}$, etc.) results in a net shift in the interferometer phase. If this phase shift changes with source mass position, it would manifest as a false acceleration signal.

Clipping: Effects from the source mass on the cavity mode are negligible because the hole in the source mass, through which the cavity mode passes, is >16 waists wide. No effect on the cavity linewidth or coupling efficiency could be observed by placing the source mass in the *near* position. However, the source mass slightly clips two MOT beams due to geometric constraints, which could lead to a difference in the radial distribution of the launched atoms. This would lead to a difference in ac Stark shifts. This problem is exacerbated by the small (600 µm) beam waist of our cavity mode. Since the beam waist is of order the size of the atom cloud, the spatial dependence of ac Stark shifts across the cloud is non-negligible. These shifts can be suppressed by k-reversal, but the cavity complicates this.

Frequency generation: The Raman frequency pair is generated by an electro-optic modulator (EOM) phase modulating the Raman laser, creating sidebands. The cesium hyperfine frequency (9 192 631 770 Hz) is just short of 25 cavity free spectral ranges (367.849 MHz). This leaves each sideband (3.6 MHz ± $\delta_{\text{Dopp}}/2\pi$) away from cavity resonance if the cavity is locked to the carrier beam. During the interferometer, we ramp the frequency difference of the Raman beams by 2.42 MHz to compensate the free fall Doppler shift. In a cavity of linewidth $\Gamma = 3.03$ MHz, this means that the third pulse (if naively locking the cavity to carrier resonance) creates a different light field within the cavity than the first pulse. Extended Data Fig. 2B illustrates these frequencies in relation to each other.

Cavity effects: Applying the cavity transfer function to the three incident beams (carrier and ±1 order sidebands from the EOM) gives intensities and phase shifts that are in general quite different from each other. Furthermore, inverting the wavevector inverts the direction of the Doppler-compensation ramp. Without careful attention, the ac Stark phases $\Delta\phi_j^{\pm}$ from pulses $k_j^{\pm}$ will not be equal (where $j = 1, 3$ indicates the pulse number in an interferometer with wavevector $k^{\pm}$). For k-reversal to effectively cancel ac Stark phases, the $\Delta\phi_j^{\pm}$ should be made approximately equal.

Finding a cavity offset: We solve this problem by offsetting the cavity resonance from the carrier such that pulses $k_j^{\pm}$ give the same ac Stark shifts. Our protocol to measure the ac Stark shift is shown in Extended Data Fig. 2A. This measurement is performed as a function of cavity offset to find offsets at which the ac Stark shifts are made to be equal.

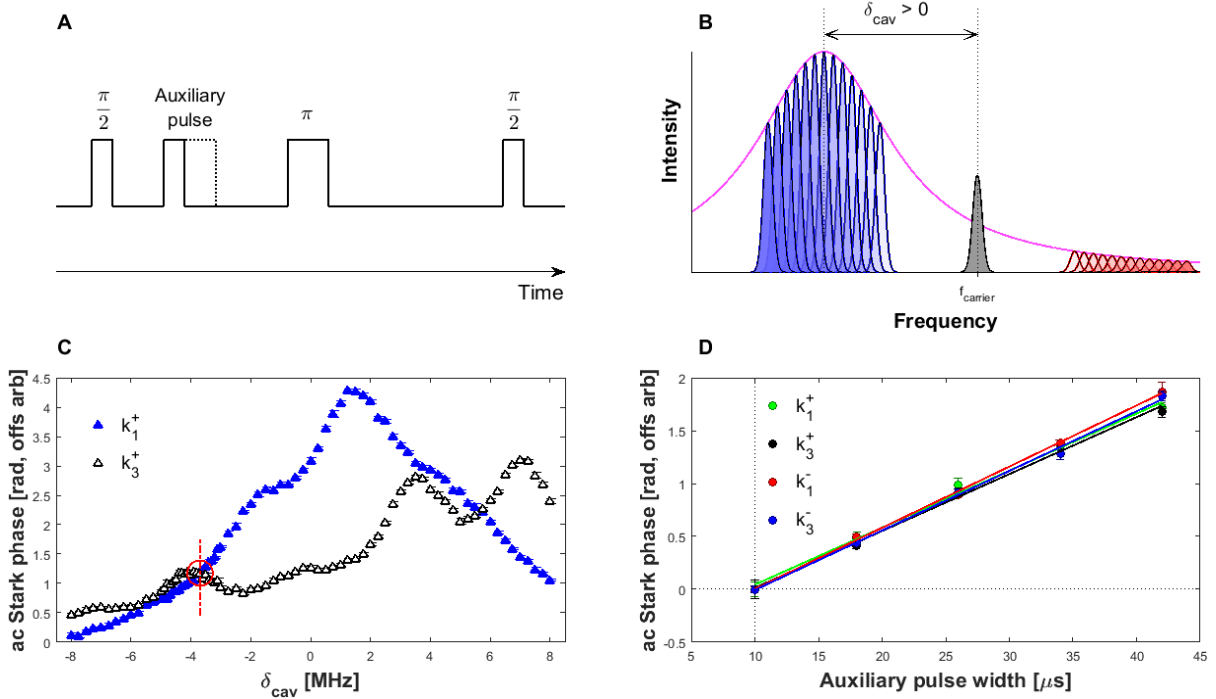

**Extended Data Figure 2 | AC Stark shifts. A)** AC Stark measurement protocol. An auxiliary pulse of variable length during a Mach-Zehnder interferometer is used to measure the ac Stark shift. The atoms are launched at a smaller velocity for this measurement. The auxiliary pulse thus does not drive transitions due to a large Doppler detuning from Raman resonance. **B)** Beams within the cavity. Cavity lineshape shown in magenta and the carrier beam shown in black. Blue (red) indicates the +(−)1 order sideband. The height of each laser lineshape indicates its intensity inside the cavity. For $k^+$, the Doppler compensation ramp moves the sidebands from transparent to opaque lineshapes. For $k^-$, the ramp moves in the opposite direction. **C)** AC Stark phase vs. cavity offset $\delta_{cav}$. The traces shown are for a single wavevector direction $k^+$ for pulses 1 and 3. The point where they intersect is the cavity offset used for $k^+$ in the actual measurement of $a_{cyl}$. **D)** AC Stark shift for the four relevant interferometer pulses. The ac Stark shift $\Delta f_j^\pm$ is given by the slope of the linear fit. The relative misalignment of the lines gives our k-reversal inefficiency $\varepsilon_{ac}$. The cavity offset used was −3.6 MHz (−4.1 MHz) for $k^+$ ($k^-$).

The phase shifts $\Delta\phi_j^+$ as a function of cavity offset from carrier resonance $\delta_{cav}$ are shown in Extended Data Fig. 2C. The red circle indicates the point where pulses $k_1^+$ and $k_3^+$ impart the same phase shifts. Operating the $k^+$ interferometer here minimizes the ac Stark phase shift ($\Delta\phi_3^+ - \Delta\phi_1^+$). We find the analogous cavity offset for $k^-$ as well. The ac Stark shifts $\Delta f_j^\pm$ are shown in Extended Data Fig. 2D. The mismatch of the $\Delta f_j^\pm$ for these four pulses, $\varepsilon_{ac} = \frac{\text{range}(\Delta f_j^\pm)}{\text{mean}(\Delta f_j^\pm)}$ is <10%.

Uncertainty: The estimated error in the acceleration measurement arising from the ac Stark shift can then be given by

$$\Delta a_{ac} \lesssim \varepsilon_{ac} \times \frac{1}{k_{eff}T^2}\{\Delta\phi_{ac}^{near} - \Delta\phi_{ac}^{far}\}, \qquad (8)$$

where $\phi_{ac}^i$ is the ac Stark phase for the interferometer when the source mass is in position $i$. We can infer the bracketed quantity from a given dataset using the measured data:

$$\{\Delta\phi_{ac}^{near} - \Delta\phi_{ac}^{far}\} = k_{eff}T^2 \times \{(a_{near}^+ - a_{far}^+) + (a_{near}^- - a_{far}^-)\}, \qquad (9)$$

where $a_i^\pm$ is the measured acceleration for wavevector $k^\pm$, with source mass position $i$. In short, we cancel source mass-dependent ac Stark shifts with inefficiency $\varepsilon_{ac}$. This varies somewhat across datasets, but a weighted average across datasets gives 8 nm/s² uncertainty.

*Vertical Alignment*

Since the measured acceleration is $\vec{k}\cdot\vec{a} = ka\cos\theta$, ensuring that $\theta = 0$ gives the true acceleration, as well as reduces sensitivity to tilt fluctuations. Around $\theta = 0$, tilt changes affect the measurement only quadratically as $a \to a \cdot (1-\theta^2)$. Toggling the source mass could introduce a systematic tilt, which could be mistaken for an acceleration signal. We stabilize the cavity mode wavevector along Earth's gravity using a feedback loop with ~1 min time constant. This is faster than the source mass toggling, but slow compared to the experimental cycle time.

An electronic bubble level (Applied Geomechanics 700-series) mounted to the vacuum chamber measures the chamber's tilt. Feedback is actuated with needle valves regulating the height of the floating optical table legs. The setpoint was determined by mapping the measured acceleration as a function of $\theta_x$ and $\theta_y$ (the tilt angles along the two axes) and finding the maximum. An example is shown in Extended Data Fig. 3.

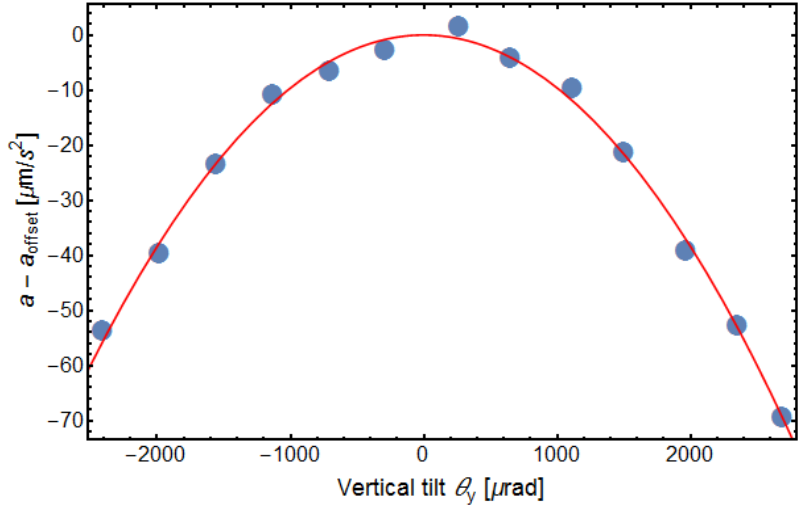

**Extended Data Figure 3 | Tilt calibration.** The measured acceleration varies $\propto \cos\theta_y \approx (1 - \theta_y^2)$ for tilt angle $\theta_y$.

The feedback is necessary, *e.g.*, to compensate for the drift of the pneumatic vibration isolation. For our datasets, the tilt data from the bubble level differ with source mass position by under 5 μrad for each dataset, corresponding a systematic effect of $< 0.2$ nm/s².

*Drifts*
Environmental effects such as tides, laser power drifts, or temperature changes can cause the measured accelerations to drift on long time scales. If we always measure {*near*, *far*} in the same order it is conceivable that a long, slow drift could cause the second position to produce a systematically different result (*e.g.*, slightly higher for a drift upwards). The effect is largely suppressed for sufficiently fast toggling of the source mass position. Remaining effects are quantified by fitting a polynomial to the acceleration measurements with source mass in the *far* position only. We then subtract this polynomial from all the data. A weighted average of the absolute value of these drift corrections across datasets results in a 3 nm/s² shift, which we quote as a drift uncertainty.

*DC Stark shift*
The source mass is electrically grounded. However, thin films of surface oxidation ~10 nm thick make form an insulating layer, allowing a voltage to build up. These films may have a dielectric strength of up to several MV/cm, allowing for surface voltages of up to 10 V. From the ground state dc polarizability of cesium, even this maximal voltage would cause a maximum acceleration of only 0.8 nm/s².

*Source mass gravity*
We model the source mass as a hollow cylinder with a wedge subtending an angle $\theta$ removed. We calculate the vertical acceleration due to gravitational attraction a test particle would feel as a function of position, and integrate this along the atomic trajectory. The characterization of the source mass dimensions reproduces the density of tungsten to within <1%. Thus, the largest source of error in the gravitational pull is the positioning. Even assuming a large positioning error of 2 mm, the average acceleration experienced by the atoms changes only by 5%, ~3 nm/s². To be even more conservative, accounting for transverse positioning, etc., we use 5 nm/s² as the uncertainty in the gravitational attraction of the source mass.

**Error budget**

Systematic uncertainties for the individual datasets are combined, weighted by the statistical uncertainties of the datasets. Extended Data Table 1 shows the resulting error budget.

**Extended Data Table 1 | Error budget**

| Quantity | Correction [nm/s$^2$] | Uncertainty [nm/s$^2$] |
|---:|:---:|:---:|
| Zeeman shift | - | 13 |
| AC Stark shift | - | 8 |
| Vertical alignment | - | 0.2 |
| Drift | - | 3 |
| DC Stark shift | - | 0.8 |
| Source mass gravity | 65 | 5 |
| Total | 65 | 16 |

**Chameleon fields**

The chameleon mechanism in relation to atom interferometry has previously been discussed in detail[9,10,22], and so will only be summarized here. It relies on an interplay between the self-interaction potential $V$ and the chameleon-matter interaction potential $V_{\text{int}}$ such that the mass of the chameleon particles increases as a function of matter density. Since the range of a force is inversely proportional to the mass of its mediating particle, the chameleon force becomes very short-ranged in typical fifth force measurements (which are often conducted in atmosphere with macroscopic objects), preventing it from being detected.

We focus here on one of the simplest, most widely-studied theories exhibiting this mechanism, where the chameleon's effective potential is given by

$$V_{\text{eff}}(\varphi) = \Lambda^4 + \frac{\Lambda^{4+n}}{\varphi^n} + \frac{\rho_m \varphi}{M} \,. \tag{10}$$

The mass of the chameleon particle is $m^2 = \partial^2 V_{\text{eff}}/\partial \varphi^2$, and therefore receives contributions from the second and third terms of equation (10). When the environmental density $\rho_m$ is large, the minimum of the effective potential is at a small value of $\varphi$, leading to a large chameleon particle mass and a correspondingly short-ranged force. Conversely, when $\rho_m$ is small, so is $m$, resulting in a long-ranged force. This effectively screens the effects of chameleon particles in dense environments.

**Symmetron fields**

A symmetron scalar field has an effective potential symmetric under $\varphi \to -\varphi$, whose low-density minima break this symmetry, and whose matter interaction restores it at high densities. A simple example is the w-shaped double well potential shown in Extended Data Fig. 4. At low densities, the field picks one of the minima, and thus breaks the symmetry. As we will show, the symmetron's interaction with a small test particle is proportional to its field value. Thus in this "asymmetric phase", with a nonzero field value, the symmetron mediates a fifth force between test masses. The simplest interaction term $V_{\text{int}}$ is a v-shaped term quadratic in $\varphi$, also shown in Extended Data Fig. 4. At high densities, this dominates, and the entire effective potential becomes v-shaped, making the symmetric value $\varphi = 0$ the minimum of the potential. In this "symmetric phase", small changes to the density no longer alter the field value, and the fifth force vanishes. Thus the symmetron "hides" its fifth force by restoring its symmetry in high-density regions.

The above statements follow from the effective symmetron potential $V_{\text{eff}}$ which, as with the chameleon, is the sum of a bare potential $V$ and an interaction potential $V_{\text{int}}$. The simplest effective symmetron potential, which we study to illustrate the mechanism, takes the form

$$V_{\text{eff}}(\varphi) = \frac{\lambda}{4!}\varphi^4 + \frac{1}{2}\left(\frac{\rho_m}{M_S^2} - \mu^2\right)\varphi^2 \,, \tag{11}$$

in which $\lambda$ is the self-coupling, $M_S$ is the matter coupling suppression scale, and $\mu$ is the bare potential mass scale. Moreover, $\lambda$ is dimensionless, while $M_S$ and $\mu$ have units of energy. This model can also be constructed in a way that is radiatively stable with well-behaved quantum corrections[33].

Extended Data Fig. 4 shows this potential for matter densities lower and higher than the characteristic density $\mu^2 M_S^2$, which is 0.23 g/cm$^3$ for the parameters shown. In regions of low density, the field minimizes the potential by choosing one of the two minima $\pm\mu\left(\frac{6}{\lambda}\right)^{1/2}$, breaking the $\varphi \to -\varphi$ symmetry. At high density, it settles at the symmetry-preserving value $\varphi = 0$.

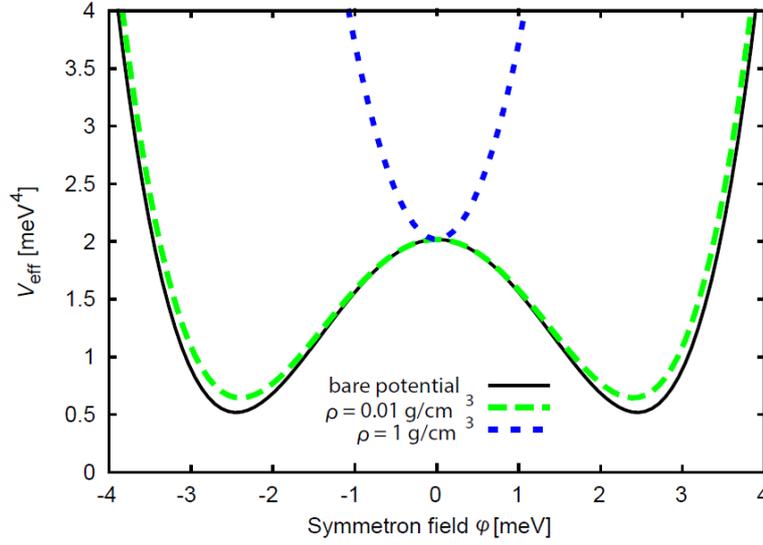

**Extended Data Figure 4 | Symmetron potentials.** At low and high densities for $\lambda = 1$, $\mu = 1$ meV, and $M_s = 1$ TeV. At low density, the field chooses one of the symmetry-breaking minima of the double-well potential. At high density, the symmetry-restoring minimum of the potential causes the effective matter coupling to vanish.

Symmetron screening is illustrated by linearizing the symmetron equation of motion[34] about a constant background field value, $\varphi = \bar{\varphi} + \delta\varphi$. In the static, non-relativistic case,

$$\nabla \delta\varphi = \frac{\bar{\varphi}}{M_S^2}\rho_m + 2\mu^2 \delta\varphi \qquad (12)$$

which is analogous to the Poisson equation for the gravitational potential but with an additional mass term and an effective matter coupling $\beta_{\text{sym}}(\varphi) = \varphi M_{\text{Pl}}/M_S^2$. In the low-density limit, this approaches $6.0\,\lambda^{-\frac{1}{2}}\left(\frac{\mu}{\text{meV}}\right)\left(\frac{M}{\text{TeV}}\right)^{-2}$. At high density, $\rho_m > \mu^2 M_S^2$, the field sits at $\varphi = 0$, and the effective matter coupling vanishes. Thus fifth forces are suppressed at high $\rho_m$.

In the general case, the field profile due to a source mass is found by solving the non-linear Poisson equation $\nabla^2 \varphi = \partial V_{\text{eff}}/\partial \varphi$ just as with chameleons. Given that source field, we can describe the acceleration of an atom using the effective coupling $\beta_{\text{sym}}(\varphi)$ and a screening parameter $\lambda_a$ which we must determine. Using the correspondence between the linearized symmetron equation and the Poisson equation for a linear test particle, we see that $\delta\varphi = 2M_{\text{Pl}}^2 M_S^{-2} \bar{\varphi}\Psi$, where $\Psi$ is the gravitational potential due to the test particle. This linear treatment breaks down when $|\delta\varphi| = \bar{\varphi}$, corresponding to $\Psi = \frac{M_S^2}{2M_{\text{Pl}}^2} = 8.4 \times 10^{-32}\left(\frac{M_S}{\text{TeV}}\right)^2$. Thus $\lambda_a$ will be nearly unity as long as the gravitational potential of the atomic nucleus is smaller than this value. Approximating the cesium nucleus as a uniform-density sphere of radius $r = 1.25\,A^{1/3}$ fm with $A = 133$, we find a gravitational potential $2.6 \times 10^{-38}$, meaning that $\lambda_a = 1$ is accurate for $M_S$ greater than 1 GeV.

Below $M_S = 1$ GeV, the atomic nucleus is partially screened. To obtain the screening factor $\lambda_a$, we divided the expression for the scalar charge[11,12] by itself in the unscreened limit. This yields a value between zero (strongly screened) and one (unscreened).

## Numerical simulations

Computing the force from the scalar field on the atom involves solving equation (5), a non-linear Poisson equation, for a matter distribution $\rho_m(\vec{x})$ that corresponds to the experimental setup. This includes the source mass, as well as the walls of the vacuum chamber. We do not compute the contribution from the atoms themselves in the calculation of $\varphi$; this effect is captured by the screening factor $\lambda_a$.

Our approach is to use a Gauss-Seidel finite-difference relaxation scheme on a three-dimensional grid that covers the entire experiment. An initial guess for the field inside the vacuum chamber is iteratively corrected until the field value converges everywhere. We have previously used this technique in the context of chameleons[22], which we have repeated for the new chameleon constraints and also extended to symmetrons. Once the field profile is known, equation (1) can be used to calculate the acceleration. We compute the average acceleration due to the scalar field as a time-weighted average over the trajectory of the atoms during the measurement.

Since the calculation is being done in near-vacuum, it is reasonable to expect the field profile to be roughly independent of $M$ (for the chameleon) and $M_S$ (for the symmetron). This is because those parameters only appear in their equation of motion along with $\rho_m$, which is very small. Extended Data Fig. 5A demonstrates this for the chameleon field, showing that the field's gradient is unchanged over five orders of magnitude in $M$.

Similarly, the vacuum value of $\varphi$ for the symmetron field is inversely proportional to the square root of $\lambda$, so we might expect $\sqrt{\lambda}\varphi$ to be independent of $\lambda$. Indeed, Extended Data Fig. 5B shows $\lambda\,\varphi\,\vec{\nabla}\varphi$ to be independent of both $M_S$ and $\lambda$ over six and ten orders of magnitude respectively. This finding greatly expedites the numerics, as only a single simulation need be performed for a given value of $\mu$.

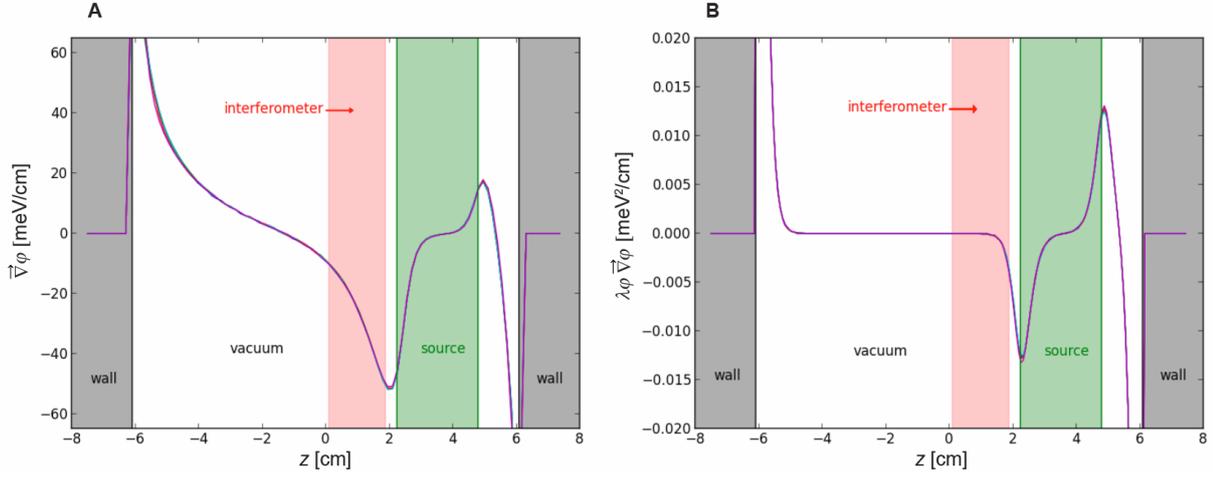

**Extended Data Figure 5 | Field profiles. A)** The gradient of the chameleon field $\vec{\nabla}\varphi$ vs. position inside the vacuum chamber, for $\Lambda = 2.4$ meV, $n = 1$, and $10^{-4}\,M_{\text{Pl}} < M < M_{\text{Pl}}$. We see that the gradient of the field, which is proportional to the chameleon acceleration, is independent of $M$. **B)** The combination $\lambda\varphi\,\vec{\nabla}\varphi$ for the symmetron field, plotted for $\mu = 10^{-1}$ meV, $10^{-3}$ GeV $< M_S < 10^3$ GeV, and $10^{-10} < \lambda < 1$. We see that the combination, which is proportional to the symmetron acceleration, is independent of $M_S$ and $\lambda$.

As with Burrage et al.[11], we find a measurable acceleration only for a relatively narrow range of $\mu$: roughly $10^{-1.5}$ meV $< \mu < 10^{-1}$ meV. In fact, the lower end of our range in $\mu$ is an order of magnitude higher than that of Burrage et al. This is because our 3D numerical code more accurately accounts for the presence of the vacuum chamber walls, which generically causes the field to vanish below a certain value of $\mu$ (or $M_S$, as seen in Fig. 3C). The upper end of $\mu$ is unchanged, and is due to the symmetron field becoming too short-ranged for the atoms to feel any appreciable force from the source mass.